\documentclass{emulateapj}
\def\stacksymbols #1#2#3#4{\def\theguybelow{#2}
        \def\verticalposition{\lower#3pt}
        \def\spacingwithinsymbol{\baselineskip0pt\lineskip#4pt}
        \mathrel{\mathpalette\intermediary#1}}
\def\intermediary #1#2{\verticalposition\vbox{\spacingwithinsymbol
        \everycr={}\tabskip0pt
        \halign{$\mathsurround0pt#1\hfil##\hfil$\crcr#2\crcr
                \theguybelow\crcr}}}

\shorttitle{Extended Dust in Elliptical Galaxies}
\shortauthors{Temi, Brighenti \& Mathews}

\begin{document}

%\title{EXTENDED, HIGHLY TRANSIENT DUSTY PLUMES IN TWO 
%ELLIPTICAL GALAXIES}
\title{{\it SPITZER} OBSERVATIONS OF TRANSIENT, 
EXTENDED DUST IN TWO 
ELLIPTICAL GALAXIES: NEW EVIDENCE OF RECENT 
FEEDBACK ENERGY RELEASE IN GALACTIC CORES}
%OF NGC 5044 AND 4636} 

\author{Pasquale Temi\altaffilmark{1,2},
Fabrizio Brighenti\altaffilmark{3,4}, William
G. Mathews\altaffilmark{3} }
%\email{ptemi@mail.arc.nasa.gov}
%\email{mathews@ucolick.org}
%\email{fabrizio.brighenti@unibo.it}

\altaffiltext{1}{Astrophysics Branch, NASA/Ames Research Center, MS
  245-6,
Moffett Field, CA 94035.}
\altaffiltext{2}{SETI Institute, Mountain View, CA 94043;
and Department of Physics and Astronomy, University of Western
Ontario,
London, ON N6A 3K7, Canada. ptemi@mail.arc.nasa.gov}
\altaffiltext{3}{University of California Observatories/Lick
  Observatory,
Board of Studies in Astronomy and Astrophysics,
University of California, Santa Cruz, CA 95064 
mathews@ucolick.org}
\altaffiltext{4}{Dipartimento di Astronomia,
Universit\`a di Bologna, via Ranzani 1, Bologna 40127, Italy 
fabrizio.brighenti@unibo.it}

\begin{abstract}
{\it Spitzer} 
observations of extended dust in two optically normal
elliptical galaxies provide a new confirmation of buoyant feedback
outflow in the hot gas atmospheres around these galaxies.  AGN
feedback energy is required to prevent wholesale cooling and star
formation in these group-centered galaxies. In NGC 5044 we observe
interstellar (presumably PAH) emission at 8 $\mu$m out to about 5
kpc. Both NGC 5044 and 4636 have extended 70 $\mu$m emission from cold
dust exceeding that expected from stellar mass loss.  The sputtering
lifetime of this extended dust in the $\sim1$keV interstellar gas,
$\sim10^7$ yrs, establishes the time when the dust first entered the
hot gas.  Evidently the extended dust originated in dusty disks or
clouds, commonly observed in elliptical galaxy cores, that were
disrupted, heated and buoyantly transported outward.  The surviving
central dust in NGC 5044 and 4636 has been disrupted into many small
filaments.  It is remarkable that the asymmetrically extended 8 $\mu$m
emission in NGC 5044 is spatially coincident with Halpha+[NII]
emission from warm gas. A calculation shows that dust-assisted cooling
in buoyant hot gas moving out from the galactic core can cool within a
few kpc in about $\sim10^7$ yrs, explaining the optical line emission
observed.  The X-ray images of both galaxies are disturbed.  All
timescales for transient activity -- restoration of equilibrium and
buoyant transport in the hot gas, dynamics of surviving dust
fragments, and dust sputtering -- are consistent with a central
release of feedback energy in both galaxies about $10^7$ yrs ago.

\end{abstract}

\keywords{galaxies: elliptical and lenticular; galaxies: ISM;
infrared: galaxies; infrared: ISM}

\section{Introduction}
                                   
In a recent survey of elliptical galaxies observed with 
the {\it Spitzer} telescope 
we found spatially extended cold interstellar dust emitting at 
70 $\mu$m in a significant fraction of our sample 
(Temi, Brighenti \& Mathews 2007).
A few unusual galaxies in our sample, having 
atypically large and massive dust lanes
that often dominate the optical image, 
probably acquired their dust and cold gas from a merger.
However, some optically normal elliptical galaxies also 
contain extended cold dust, such as the two galaxies we 
discuss here, NGC 5044 and 4636.
Evidently, the source of this dust is internal, not from mergers. 
Since this dust is thought to be in direct contact with 
the hot, virialized interstellar gas ($kT \sim 1$ keV), 
it has a short lifetime ($\sim 10^7$ yrs)  
to sputtering destruction by thermal ions. 
Consequently, this dust is a spatial tracer of extremely transient 
events that recently occurred on kpc scales. 
%Here we present images and a detailed 
%analysis of two nearby elliptical galaxies with extended dust, 
%NGC 5044 and NGC 4636. 
Both galaxies are extended at 70 $\mu$m, but NGC 5044 
is also extended at 8 $\mu$m in a manner similar to the 
highly asymmetric optical line emission from warm gas 
in this galaxy. 
The temperature of the warm gas is maintained by photoionization 
by UV radiation from hot post-asymptotic giant branch (AGB) 
stars and radiation 
of optical line emission; the dust is heated by starlight and 
hot thermal electrons and cooled by infrared emission. 

As we proposed in Temi, Brighenti \& Mathews (2007), 
current evidence is consistent with an internal origin for this dust, 
and we hypothesize that dust has been buoyantly transported from the 
galactic cores out to several kpc following a feedback heating 
event. 
In both galaxies 
small, disorganized fragments of optically absorbing dusty gas 
are visible within the central $\sim$100 parsecs, 
further evidence of a recent central energy release.
In addition, {\it Chandra} X-ray images reveal that 
the extended hot interstellar gas in 
both galaxies is undergoing short-term gasdynamical activity.
All this information converges to confirm the 
transient state of affairs in these galaxies. 
In particular, the association of interstellar PAH emission 
(from polycyclic aromatic hydrocarbon molecules) and warm gas 
($T \sim 10^4$ K) in NGC 5044 indicates that we may be 
viewing this galaxy at a rare moment immediately following 
a release of energy near the central black hole. 
This type of central feedback 
energy release, thought to be caused by 
the accretion of a small mass of gas into the
central black hole, has been  
proposed as a likely solution of 
the so-called cooling flow problem. 
In the following we adopt distances of 15 Mpc 
($1^{\prime} = 4.36$ kpc) for NGC 4636 
and 33 Mpc ($1^{\prime} = 9.60$ kpc) for NGC 5044 
(Tonry et al. 2001).

\section{Images of NGC 5044 and NGC 4636}

Figures 1 and 2 show {\it Spitzer} images of NCC 5044 
and NGC 4636 at  
160, 70, 24, and 8 $\mu$m. 
The 8-4.5 $\mu$m images are 
differences of {\it Spitzer} images at 8.0 and 4.5 $\mu$m. 
Details of the acquisition and reduction of far-infrared 
data at 24, 70 and 160 $\mu$m are described in 
Temi, Brighenti \& Mathews (2007). 
These data were obtained with the Multiband Imager Photometer (MIPS)
(Rieke et al. 2004) on the {\it Spitzer} Space Telescope.
Both NGC 5044 and 4636 are detected at 160 $\mu$m, but neither
image can be distinguished from a point source.
However, as we illustrate in Figure 10 of
Temi, Brighenti \& Mathews (2007), at
70 $\mu$m both NGC 5044 and 4636 are
clearly extended beyond the MIPS point response function. 

Mid-infrared data at 3.6, 4.5, 5.8 and 8.0 $\mu$m 
for NGC 4636 were obtained using the Infrared Array 
Camera (IRAC) onboard {\it Spitzer} 
under the guaranteed time program (PID 69) 
led by PI G. Fazio (Fazio et al. 2004). 
NGC 4636 was observed for 60 seconds on each of the 
four IRAC wavelength channels. 
NGC 5044 was observed during an IRAC calibration program
(PID 1160, PI: W. T. Reach). 
Images of NGC 5044 at each of the four IRAC bandpasses were 
recorded at 5 different positions slightly
offset from the galaxy center. 
Observations at each position were acquired using
a five position gaussian dither parameter with an integration time
of 12 seconds per frame. 
Thus, the total on target integration time
at IRAC wavelengths was 300 seconds.
For all wavelengths and with both MIPS and IRAC, full coverage
imaging was obtained for all observations with additional sky
coverage to properly evaluate the background emission. 
We used the Basic Calibrated 
Data (BCD) products from the Spitzer
Science pipeline (version 13.2) to construct mosaic
images when necessary.
Pipeline reduction and post-BCD processing
using the MOPEX software package provide all necessary steps
to process individual frames: dark subtraction,
flat-fielding, mux-bleed correction, flux calibration, correction
of focal plane geometrical distortion, and cosmic ray rejection.

Also shown in Figures 1 and 2 are 
images of the H$\alpha$+[NII] emission 
from the warm gas in NGC 5044 (isophotes) 
and 4636 (gray scale).
The panels at the lower right in both figures illustrate 
isophotes of the X-ray images of 
the same central regions of both 
galaxies, produced from data in the {\it Chandra} 
archives, superimposed on optical images from the 
Digital Sky Survey.

The H$\alpha$+[NII] image of NGC 5044 is quite unusual, 
with extensions 
toward the East, North and (especially) to the South. 
These warm gas features are obviously 
unrelated to the underlying stellar distribution. 
The H$\alpha$+[NII] image in NGC 4636 is more 
compact, but nevertheless 
exhibits irregularities in the outer regions. 
The radial velocities of the warm gas 
in both galaxies determined 
from slit spectra show highly irregular 
activity of $\pm \sim 150$ km s$^{-1}$ on scales of 
$\sim 1$ kpc that are unrelated to the bulk motion of 
the stars 
(Caon et al. 2000).
These quasi-random velocities are consistent with 
the disordered spatial appearance of the warm gas.

\section{Excess Dust in NGC 5044 and 4636 and its Origin}

Recent studies of mid-infrared {\it Spitzer} IRS spectra 
of elliptical galaxies indicate that most of the emission 
shortward of 8 $\mu$m is photospheric, while emission 
in the 8-20 $\mu$m region is from circumstellar dust around 
mass-losing red giant stars 
(Bregman, Temi \& Bregman 2006; Bressan et al. 2006).
In addition,  
Temi, Brighenti \& Mathews (2007) find a tight correlation 
between fluxes at 24 $\mu$m from {\it Spitzer} MIPS data 
and optical B-band fluxes, 
suggesting that circumstellar dust also dominates at this 
wavelength.
But MIPS fluxes from elliptical galaxies 
at 70 and 160 $\mu$m are uncorrelated with optical fluxes from 
starlight, so we regard this colder dust as truly 
interstellar. 
It is remarkable that the 
fluxes at 70 and 160 $\mu$m vary by factors of 30-100 
among galaxies with similar optical fluxes.
Some elliptical galaxies with large 
70 and 160 $\mu$m fluxes contain very extended, multi-kpc 
disks of cold, dusty gas consistent with 
the strength and spatial extent of the 70 $\mu$m image. 
Evidently, these galaxies have experienced an unusual major 
merger with a gas-rich spiral.
However, a few optically normal elliptical galaxies, 
including NGC 5044 and 4636, 
also show unexpected large 70 and 160 $\mu$m fluxes with 
70 $\mu$m emission extended 
well beyond the MIPS point response function. 

Because the 8-24 $\mu$m spectrum of elliptical galaxies is 
dominated by circumstellar dust in outflowing stellar 
winds, some internally produced dust is expected when 
this dust moves into the interstellar environment, 
where it is heated by diffuse starlight 
to much lower temperatures
and radiates in the far-infrared. 
In Temi, Brighenti \& Mathews (2007) we describe a 
simple steady state model in which interstellar dust 
is continuously produced by normal mass loss from 
a single population of evolving stars and continuously 
destroyed by ion sputtering. 
A far-infrared spectral energy distribution (SED) can 
be calculated assuming a power-law initial 
grain size distribution 
$\propto a^{-3.5}$ in $0 < a < a_{max}$. 
This grain size distribution is normalized so 
that $\delta = (1/150)z(r)$ is the ratio of dust to 
gas mass in the circumstellar outflows where 
$z(r) \approx (r/R_e)^{-0.207}$ is a typical 
radial variation of the metal abundance (in solar units) 
in elliptical galaxies (Arimoto et al. 1997).
We assume amorphous silicate grains and compute the 
temperature of each dust grain. 
The grain production rate is proportional to the local 
stellar density and the specific stellar mass loss rate 
for an old stellar population 
$\alpha_* = {\dot M}_*/M_* 
= 4.7 \times 10^{-20}$ s$^{-1}$ (Mathews 1989).
The sputtering rate of the grains is determined by the local 
density and temperature in the hot gas as determined 
from X-ray observations. 
The total galactic SED is found by integrating the dust emission over 
the steady state dust size distribution and galactic radius. 
We show in Figure 11 
of Temi, Brighenti \& Mathews (2007) that the 
far-infrared SED at 70 and 160 $\mu$m observed in at least a 
few galaxies -- such as NGC 1399, 1404 and 4472 -- 
can be fully explained by this type of normal stellar mass loss. 
Elliptical galaxies that agree with our steady state 
dust model are among those with the lowest 
70 and 160 $\mu$m luminosities.

However, as illustrated in Figure 3, 
the 70 and 160 $\mu$m fluxes of both NGC 5044 and 4636 
are far in excess of our predicted steady state SED. 
(The 24 $\mu$m flux shown in Figure 3 is circumstellar 
and is unrelated to our SED model.)
The mass and spatial extent 
of the ``excess'' interstellar dust in these 
galaxies can be estimated using our steady state model
but artificially increasing the stellar 
mass loss rate $\alpha_*$ within radius $r_{ex}$. 
Increasing $\alpha_*$ by some factor $f_{\alpha}$ 
increases the mass of interstellar 
dust and $r_{ex}$ can be adjusted until 
the mean grain temperature approximately fits 
the 70 $\mu$m/160 $\mu$m flux ratio observed in each galaxy. 
Interstellar dust is heated by starlight and 
by inelastic collisions with thermal electrons. 
Since the mean intensity of starlight and the 
electron density both decrease
with galactic radius, the dust temperature 
$T_d$ also decreases, so $r_{ex}$ can be adjusted to change 
mean dust grain temperature $\langle T_d \rangle$ 
and therefore the 70 $\mu$m/160 $\mu$m flux ratio. 
These values of $f_{\alpha}$, $r_{ex}$ and the 
excess dust mass estimated from them are only approximations,
since it is unlikely that the radial distribution of the 
excess non-stellar dust exactly follows the stellar density 
profile, but with a sharp truncation at $r_{ex}$. 

Each panel of Figure 3 shows three combinations of 
$f_{\alpha}$ and $r_{ex}$, holding $a_{max} = 0.3 \mu$m constant.
The best-fitting SED for the extra dust in NGC 5044 and 
4636, shown by dotted lines in both panels, corresponds 
respectively to a mass of excess dust of 
$1.5 \times 10^5$  and $1.1 \times 10^5$ $M_{\odot}$.
The physical size of this dust, $r_{ex} \sim 4-5$ kpc, 
is comparable to or 
exceeds the 70 $\mu$m point response function for both galaxies.
We propose below that $r_{ex}$ is limited by the sputtering time 
in buoyant dusty gas flowing from galaxy cores.

We argue in Temi, Brighenti \& Mathews (2007) that the 
incidence of (otherwise normal) elliptical galaxies with 
excess dust emission at $70 \mu$m  
exceeds the fraction of ellipticals that are likely 
to have experienced 
mergers with dusty gas-rich (dwarf) galaxies within 
a typical dust sputtering time $t_{sput} \approx 10^7$ yrs, 
evaluated a few kiloparsecs from the galactic center.
In addition, 
since $t_{sput}$ is somewhat less than the typical
free fall time from several effective radii, 
if the extended dust were a result of merging, 
ellipticals having far-infrared excesses should contain 
evidence of the merging galaxies within their 
optical images, but this is not observed.
The mean stellar ages of elliptical galaxies can be 
measured from the strength of Balmer line absorption features.
Annibali et al. (2006) find a stellar age of 
13.5 $\pm$ 3.6 Gyrs for NGC 4636 but the age for 
NGC 5044 (14.2 $\pm$ 10. Gyrs) is indeterminate because 
the Balmer absorption is masked by the much stronger 
interstellar Balmer emission in this galaxy. 
The optical colors of both galaxies also seem 
normal, $B-V \approx 0.97-1.06$ (Poulain 1988; 
Poulain \& Nieto 1994).
Finally, the stellar velocity dispersion and rotation are largely 
normal for both galaxies{\footnotemark[1]}
\footnotetext[1]{
NGC 4636 does have a few minor peculiarities. 
Slit spectra of 
Caon, Macchetto \& Pastoriza (2000) indicate that the 
mean stellar radial velocity 
has a broad central {\it minimum} at position angle 117$^o$, but not 
in other directions. 
Also, for its luminosity and central stellar velocity dispersion 
NCC 4636 has an unusually low near-IR stellar surface brightness, 
causing it to deviate from the fundamental plane for galaxy cores 
(Ravindranath, et al. 2001).
}
, with no 
kinematical indication of 
an alien merging stellar system 
(Caon, Macchetto, \& Pastoriza, 2000).  
We conclude there is no optical evidence in either galaxy of 
a merger event during the previous $\sim 10^8$ yrs.

In Temi et al. (2003) we first discovered 
excess dust at 60, 90 and 180 $\mu$m in NGC 4636 
from observations with the {\it Infrared Space 
Observatory} (ISO) and proposed that the short-lived 
dust resulted from a recent merger. 
However, now that the age of the stars in 
NGC 4636 has been found to 
be very old and the dust is known to be extended, 
our earlier merger explanation is untenable. 
Instead, as discussed in Temi, Brighenti \& Mathews 
(2007), we now propose that the extended excess 
dust seen in both NGC 4636 and 5044 and other similar 
galaxies has been buoyantly
transported from the galactic core to several kpc 
following a heating event near the central black hole. 

Galactic cores are a natural source of dust. 
Mathews \& Brighenti (2003) showed that 
the $\sim100-300$kpc-sized nuclear disks
that are often observed in the nuclei of elliptical galaxies 
(e.g. Lauer et al. 2005) 
can result from stellar mass loss within 
the innermost $\sim$1 kpc 
at a rate ${\dot M}_1 \approx \alpha_* M_1 
\approx 0.1$ $M_{\odot}$ yr$^{-1}$, 
where $M_1 \sim 5 \times 10^{10}$ $M_{\odot}$ is the 
stellar mass in the central kpc. 

In the Appendix of Temi et al. (2007) we showed that the excess 
far-infrared emission at 70 and 160 $\mu$m 
cannot arise from these small nuclear dust disks/clouds but must 
come from dust distributed 5-10 kpc into the hot 
gas.{\footnotemark[2]}
\footnotetext[2]{
Cold dust inside dense, optically thick nuclear disks/clouds is
self-shielded and cannot be heated by the diffuse stellar UV radiation
from hot post-AGB stars.  
The apparent longevity of these optically thick
clouds ($\sim 10^7$ yrs) 
suggests that heating by thermal electrons in the
adjacent hot gas is also suppressed.  
Evidently magnetic fields
(tangent to the surface of the spinning disks) are sufficient to
restrict thermal conduction.  
However, when the dusty gas is heated
and dispersed by an AGN explosion, some of the dust is directly
exposed to diffuse galactic UV and hot thermal electrons that heat the
dust to radiate much more efficiently at infrared wavelengths.  
But when the same mass of
dust is contained in dense, self-shielding central disks, its infrared
luminosity is unobservably low.}
We hypothesized that the spatially extended 
dust has its origin in the nuclear disks/clouds that 
are disrupted and buoyantly transported to large radii
in the hot gas as a result of intermittent AGN-type heating 
in the galactic core. 
The mass of excess extended dust, $\sim 10^5$ $M_{\odot}$,  
is consistent with this. 
%New dusty disks form after the older ones are destroyed.
 
Consider for example the time $t_{disk}$ required to  
form an optically obscuring dusty disk of radius 
$r \sim 300$ pc. 
To block starlight, the column density 
in the disk must exceed about 
$N \sim 10^{21}N_{21}$ cm$^{-2}$, assuming solar abundance.
So the total mass of an optically dark disk of radius $r$
is at least $M_{disk} \approx [5\mu/(2+\mu)] m_p N \pi r^2 
\approx 3 \times 10^7 N_{21} r_{300}^2$ $M_{\odot}$ where
$r_{300}$ is the disk radius in units of 300 parsecs 
and $\mu = 0.61$ is the mean molecular weight.
[This disk mass is consistent with the upper limit to the 
HI mass in NGC 4636 of $1.1 \times 10^8$ $M_{\odot}$
established by Krishna Kumar \& Thonnard (1983).]
The time for stellar mass loss 
(within the central kpc) to form such an obscuring disk, 
$t_{disk} \approx M_{disk}/{\dot M}_1$, 
is only $t_{disk} \approx 3 \times 10^7 N_{21} r_{300}^2$ yrs,
similar to the buoyant rise time in the hot gas.
For a dust to gas mass ratio $\delta \sim 0.01$, these disks 
should contain about 
$3 \times 10^5$ $M_{\odot}$ of dust, which is very similar 
to the dust we observe in NGC 5044 and 4636 
in excess of our steady state model.

To support this hypothesis further, 
we show in Figure 4 {\it Hubble Space Telescope} images 
of the central $\sim 1-2$ kpc in NGC 5044 and 4636. 
Small, chaotically arranged dusty fragments are visible 
against the stellar background in both galaxies. 
The dense gas in these fragments is clearly in a highly 
transient state, dynamically 
orbiting in the galactic potential out to a few kpc 
where the freefall time is $\sim 10^7$ years. 
It is very unlikely that 
these relatively large optically obscuring clouds 
can be interpreted as infalling dust just 
after being ejected from stars. 
The radius of individual optically dark fragments in Figure 4 is about 
10 parsecs, which translates to 
a dust mass $\sim 3 \times 10^4$ $M_{\odot}$ 
far in excess of that from a single star 
or from a hypothetical agglomeration of several stellar outflows
during the short freefall time.
We propose instead that these dusty regions are fragments of 
a central, more organized disk-like structure that was disrupted 
about $10^7$ yrs ago by an AGN-related energy release. 
While it may seem unlikely that dense fragments can 
result from an AGN energy release and be 
accelerated outwards in this quasi-coherent manner, 
we appeal to studies of Seyfert galaxies where similar large 
and dense clouds are accelerated in just 
this manner by non-thermal, radio-emitting outflows 
(e.g. Whittle, et al. 2005). 

As soon as a dusty disks are destroyed by a significant release of 
AGN energy, another disk begins to form.
The fragments visible in Figure 4 indicate that 
some fraction of the disk received kinetic energy but 
was not irreversibly heated by the explosive AGN event.
Therefore the next generation disks in these galaxies will 
be formed out of these denser remnants of the previous disk 
and new dust that is continuously ejected from red giant stars 
in the galactic core. 
At present it is impossible to know what fraction of 
disk material is heated to supervirial (buoyant) temperatures and 
what fraction remains in the fragments visible in 
Figure 4. 
Nor is it clear at present what limits the maximum 
amount of dusty cold gas that resides in central disks/clouds.

The only signs of current nuclear activity in NGC 5044 and 4636 are
small, weak non-thermal GHz nuclear radio sources.
There is no
evidence of disk or Bondi type accretion.
For example, from Chandra
observations of NGC 4636 
Loewenstein et al. (2001) determined that the accretion
onto the $8 \times 10^7$ $M_{\odot}$ central massive black hole 
in this galaxy 
is less than $3 \times 10^{-8}L_{edd}$ and less than $6 \times
10^{-4}$ of the spherical Bondi accretion rate.
The general absence
of mass accretion in supermassive nucelear black holes in elliptical
galaxies is consistent
with models in which most of their mass was acquired during a
short-lived early evolutionary phase (e.g. Hopkins, Narayan, \&
Hernquist 2006).
The AGN-type energy releases in NGC 5044 and 4636
required in our interpretation must also be sufficiently short-lived
or radiatively unremarkable.

\section{PAH Emission from NGC 5044}

Perhaps the most dramatic illustration of extended 
dust in NGC 5044 is the difference image between IRAC wavelengths 
8 and 4.5 $\mu$m shown in Figure 1 and its 
striking similarity to that of the warm gas emission 
in H$\alpha$+[NII]. 
The 8-4.5 $\mu$m difference image is constructed by 
subtracting a properly normalized 4.5 $\mu$m image 
from the 8 $\mu$m image. 
Emission at 4.5 $\mu$m is due largely to stellar photospheres 
while both photospheric and circumstellar emission 
contribute at 8 $\mu$m -- 
but the 8 $\mu$m image may also contain interstellar emission 
from hot dust emission or the strong 7.7 $\mu$m 
PAH emission feature. 
When the two images are subtracted, both the 
stellar and circumstellar emission are canceled, 
and only emission from the faint interstellar dust emission remains. 
Our subtraction process in NGC 5044 was based on visual examination 
of the surface brightness $\Sigma$ in the difference image 
$\Sigma_{8.0} - F \Sigma_{4.5}$ where the subscripts refer 
to the IRAC wavelengths and $F$ is a variable parameter. 
As $F$ was varied, the stellar surface brightness of the difference 
image of the stellar emission passes through a cancellation when 
$F = \Sigma_{8.0}/\Sigma_{4.5}$. 
In this process 
particular attention was paid to the 
residual stellar brightness in the 
region along the H$\alpha$ plume about 3-5 kpc south of 
the galactic center. 
The value of $F$ at which the stellar contributions at 
8.0 and 4.5 microns outside the plume region 
are equalized depends on the galactic radius in NGC 5044. 
The 8-4.5 $\mu$m image of NGC 5044 shown in Figure 1 corresponds to 
$F \approx 0.52$ where the stellar emission near the plume 
canceled and the faint residual interstellar features 
are revealed. 
Our value of $F$ is almost identical to $F = 0.526$ which is 
the flux ratio 
$F_{8.0}/F_{4.5}$ of a 10 Gyr old solar abundance 
single stellar population estimated by Piovan et al. (2003), 
who include both photospheric and circumstellar infrared emission. 
In our difference image of NGC 5044 
the interstellar emission exhibits the 
same extended features toward the north and south that are so 
prominent in the H$\alpha$+[NII] isophotes. 
We estimate that the total luminosity of the 8 $\mu$m 
component is about 4 percent of the total interstellar 
far-infrared luminosity longward of 24 $\mu$m, 
this is similar 
to the PAH emission fraction observed in Milky Way sources.

The residual interstellar feature in NGC 5044 could be due 
to the strong PAH
emission band at 7.7 $\mu$m or to hot dust emission from small, 
stochastically heated dust grains, in some proportion.
Dust with temperatures approaching $10^3$ K could 
be heated by an intense beam of optical-UV radiation from 
the galactic center along the direction of the plume. 
However, to our knowledge normal elliptical galaxies 
with abnormally UV-bright nuclei, when viewed along such 
beams, are not commonly observed. 
In principle, thermal excursions of 
small grains recently heated by individual photons or electrons could 
explain the diffuse 8 $\mu$m emission. 
For example, this process is thought to explain the 
diffuse near-infrared dust emission often observed in 
planetary nebulae (e.g. Phillips \& Ramos-Larios 2006).
During the short-lived 
stages as the dusty hot gas in the southerly 
plume of NGC 5044 cools, the 
electron density increases and stochastic 
heating of small grains by electron encounters may increase. 
However, in 
situations where stochastically heated dust emission dominates, 
the infrared continuum generally increases from 8 toward 
longer wavelengths since a larger fraction of small grains can 
be momentarily heated to lower temperatures that radiate 
at longer wavelengths 
(e.g. Draine \& Anderson 1985).
But this increase is not observed in spectra 
of normal elliptical galaxies taken 
with the {\it Spitzer} Infrared Spectrograph 
(IRS) (e.g. Bregman, Temi \& Bregman 2006). 
Our 24 $\mu$m image of NGC 5044 
(not shown in Figure 1) shows no evidence 
of asymmetrically extended emission similar to the optical 
line emission.

If the extended 8-4.5 $\mu$m feature 
visible to the south in Figure 1 were due 
to transient thermal excursions of small grains, 
this feature would be even more 
pronounced in a 24-4.5 $\mu$m difference image.
To explore this possibility, we spatially 
degraded the 4.5 $\mu$m image 
to the PSF of the 24 $\mu$m image and varied the flux ratio 
to cancel the stellar and circumstellar emission 
at both wavelengths.
We see no non-stellar residual interstellar enhancement 
in the 24-4.5 $\mu$m image of NGC 5044 
that is geometrically similar to the plume extension visible 
in optical emission and the 8.0-4.5 $\mu$ difference image. 
The absence of extended emission in the 24-4.5 $\mu$m image 
of NGC 5044  
is therefore a strong argument for PAH emission at 8 $\mu$m 
and for the relative unimportance of thermal excursions 
in the mid-infrared dust SED in this galaxy. 
An IRS spectrum of NGC 5044 has been taken but it is not 
currently available in the {\it Spitzer} archive, 
although the extended interstellar emission in Figure 1 
may be too faint to be 
easily visible as PAH emission features in the spectrum.

In the following discussion we adopt a PAH interpretation of 
the 8 $\mu$m interstellar features in NGC 5044, but the 
precise identification of this dust-related interstellar emission 
is not central to our main arguments. 
It is unclear if the PAH is located in the warm gas or is 
in the hot interstellar gas where its expected lifetime 
is short but its modes of excitation 
and emission might be more easily excited.

Evidence of PAH emission in {\it Spitzer} IRS spectra 
of elliptical or other early type galaxies is rare and 
has only been found in a few unusual early-type galaxies 
that have starburst activity in massive cold disks: 
NGC 4550, NGC 4435 (Bressan et al. 2006; 
both are SB0 galaxies) and 
NGC 4697 (Bregman, Bregman \& Temi 2006; 
an E6 galaxy with a dust lane that may be an S0).

It is unclear at present if the extended 
70 $\mu$m emission also shows the same feature to the 
south of NGC 5044. 
When the 8-4.5 $\mu$m feature is convolved 
with the PRF of the 70 $\mu$ filter, this southern 
extension largely disappears.

The very elongated H$\alpha$+[NII] image in NGC 5044
is unusual -- perhaps the only other elliptical galaxy
that is known to have a similar feature
(unrelated to large disks of cold gas)
is NGC 5813 (Goudfrooij et al. 1994).
In normal galaxies without extended dust or warm gas
features, all the warm gas is expected to result from
stellar mass loss and is generally visible only
within the central kpc.
Gas ejected from evolving stars at a rate $\alpha_*M_*$
(where $\alpha_* = 4.7 \times 10^{-20}$ s$^{-1}$)
cannot accumulate in
the warm gas phase indefinitely
since otherwise after
$\sim 10^5$ yrs the H$\alpha$+[NII]
flux would exceed values commonly observed in most
elliptical galaxies.
We conclude therefore that the warm line-emitting
gas thermally merges 
with the hot ($T \sim 10^7$ K)
interstellar gas in this time
(Mathews \& Brighenti 1999).
Therefore the extended warm gas observed in NGC 5044
is very transitory, although the PAHs and warm gas
may be continuously supplied for
a longer time $\sim 10^7$ years
as gas cools in the buoyant dusty plume.

\section{Buoyant Dust Outflow in NGC 5044}

Two questions come to mind about buoyant dust transport.
Can dust grains survive sputtering destruction during 
their buoyant journey out to $\sim 5$ kpc, the approximate 
extent of the H$\alpha$+[NII] and PAH emission features 
in NGC 5044? 
Can the dust also cool (some of) the buoyant gas to create the 
warm gas that emits H$\alpha$+[NII]?
To answer these questions we consider a simple model 
in which gas is assumed to be heated 
at some small radius $r_0$ in NGC 5044 and 
remains in hydrostatic equilibrium as it 
moves buoyantly out at velocity $v$.
For simplicity we assume $v$ is constant since 
the buoyant velocity depends on 
the size of the buoyant region and other uncertain parameters. 
After the gas is heated at radius $r_0$ it has an electron density 
$n$ that is lower 
by a factor $\Delta < 1$ than the local ambient density 
of the hot gas, i.e. $n(r_0) = \Delta {\hat n}(r_0)$.
In this Section variables with hats refer to the (unheated) ambient 
hot gas in NGC 5044.
The density and temperature profiles in NGC 5044, 
found from X-ray observations (i.e. Buote et al. 
2003), can be approximated with the following 
expressions:
\begin{equation}
{\hat n} = {0.065 \over r_{kpc}^{1/2}[1 + (r_{kpc}/9.5)^{1.05}]}
+ {0.002 \over [1 + (r_{kpc}/30.5)^{3/2}]}~~~{\rm cm}^{-3}
\end{equation}
and 
\begin{displaymath}
{\hat T} = {10^7  \over 
[F(r_{kpc})^{-3} + G(r_{kpc})^{-3}]^{1/3} }
\end{displaymath}
\begin{equation}
~~~~- 0.12 e^{-(r_{kpc}/1.5)} + 3 \times 10^6~~{\rm K}
\end{equation}
where
\begin{equation}
F(r_{kpc}) = 0.85 + 0.95(r_{kpc}/30)^2
\end{equation}
and 
\begin{equation}
G(r_{kpc}) = 1.64 (r_{kpc}/30)^{-0.10}.
\end{equation}

For simplicity we assume that the grains have a single 
initial radius $a(r_0)$ and, as they move buoyantly outward, 
are sputtered at a rate given by
\begin{equation}
{da \over dr} = - { 3.2 \times 10^{-18} n_p \over v}
\left[ 1 + \left( {2 \times 10^6{\rm K} \over T}\right)^{2.5}
\right]^{-1} ~~~ {\rm cm~cm}^{-1}
\end{equation}
(Tsai \& Mathews 1995) 
where $n_p = 0.83n_e$ is the proton density.
As the buoyant gas flows out in local pressure equilibrium, 
$P = {\hat P}$, 
its temperature follows an adiabat 
($T \propto P^{(\gamma - 1)/\gamma}$ with $\gamma = 5/3$) 
modified by  radiative losses, 
\begin{equation}
{dT \over dr} = {2T \over 5{\hat P}}
\left[  {d {\hat P} \over dr}
- {1 \over v} 
\left({\mu {\hat P} \over k T} \right)^2 (\Lambda_{gas}
+ \Lambda_{dust})\right].
\end{equation}
The first term in square brackets describes the 
adiabatic cooling of the buoyant gas as it expands, 
continuously 
adjusting to the decreasing ambient pressure as it moves out
in the hot gas atmosphere of NGC 5044.
%Local pressure equilibrium is assumed, $nT = {\hat n}{\hat T}$.
The second term is the intrinsic cooling that occurs 
in the buoyant gas due to standard radiative losses
with coefficient $\Lambda_{gas}(T)$ 
(e.g. Sutherland \& Dopita 1993) and due to the 
absorption of thermal electrons by inelastic collisions
with the grains,
\begin{equation}
\Lambda_{dust} = {9 \over 8} { (2+\mu) \over 5 \mu}
{\delta m_p \over \rho_g}
\left( 8 \over \pi m_e \right)^{1/2} { (kT)^{3/2} \over a}
~~~{\rm ergs}~{\rm cm}^3~{\rm s}^{-1}
\end{equation}
(Mathews \& Brighenti 2003).
We assume a uniform  
dust mass to gas mass ratio of $\delta = 0.015$ in the initial gas, 
similar to that in Milky Way gas. 
The mass density of the grain material $\rho_g = 3.3$ g cm$^{-3}$ 
is appropriate for amorphous silicates. 

The solution to these equations shown in Figure 5 
corresponds to the following (somewhat arbitrary) parameters:
$r_0 = 200$ pc, $\Delta = 0.15$, $a(r_0) = 1\mu$m, 
and $v = 400$ km s$^{-1}$.
The upper panel of Figure 5 shows Chandra observations of 
the hot and cold temperature phases in NGC 5044 
(Buote et al. 2003). 
The solid lines in the upper two panels 
show the temperature and density profiles of 
the cooler, dense phase ${\hat T} \equiv T_c(r)$ and the 
long-dashed lines show the buoyant evolution of dust-free
gas heated by a factor $\Delta^{-1}$ at $r_0$. 
The short-dashed lines in Figure 5 show the trajectory 
of the same heated gas but with a uniform admixture of dust 
at a fraction $\delta = 0.015$ of the gas mass. 
It is seen that the buoyant gas does in fact cool 
to low temperatures ($T \ll {\hat T}$) 
at $r \approx 5$ kpc similar 
to the extent of the southern warm gas feature observed in 
NGC 5044.
The dust-assisted cooling occurs in $\sim 10^7$ yrs even 
though the buoyant gas was initially hotter than the ambient 
gas ${\hat T}$; the bottom panel in Figure 5 shows that 
dust cooling dominates at every radius in the dusty buoyant gas. 
Evidently, the PAH emission observed comes from residual 
dust fragments diminished by sputtering. 
Perhaps the least satisfactory aspect of this approximate
calculation is the rather high value of the 
buoyant velocity required to reach $\sim 5$ kpc
in $\sim 10^7$ yrs, 
$v = 400$ km s$^{-1}$, 
which nearly approaches the 
sound speed $c_s = (\gamma k {\hat T} / \mu m_p)^{1/2}
= 480 ({\hat T}/10^7~{\rm K})^{1/2}$ km s$^{-1}$.
This suggests that the buoyantly heated gas may have received 
momentum from a jet emerging from the central black hole. 

Considering the many uncertainties 
involved, it seems plausible from Figure 5 that the peculiar 
asymmetric extensions of warm gas in NGC 5044 can be formed 
by dust-assisted cooling of buoyant flow. 
Buoyant gas that was heated less at $r_0$ 
(i.e. $\Delta > 0.15$) or which contains a locally higher 
dust density will cool 
at smaller radius, explaining the full extent of 
the warm gas feature in NGC 5044.
The enhanced soft X-ray emission extending southward in 
the general direction of the plume (Figure 1)
could be due to enhanced emission from cooling 
(and therefore denser) hot gas. 
Indeed, we note that the cooling in Figure 5 occurs on a timescale 
$\sim 10^7$ yrs consistent with other transient 
events in the hot gas in 
both NGC 5044 and 4636 which we discuss below. 

\section{Central Optical and X-ray Cavity in NGC 4636}

Optical line emission from warm gas in the central 
$\sim1$ kpc of NGC 4636 and thermal 
X-ray emission from the hot gas both have a 
small cavity slightly displaced from the galactic center
{\footnotemark[3]} (Figure 6).
\footnotetext[3]{
The cavity is less visible in the emission line image 
shown in Figure 2 probably because of 
the gray-scale intensities adopted by Caon et al. (2000).
}
Demoulin-Ulrich et al. (1984), Buson et al. (1994) and
Zeilinger et al. (1996), all noticed
a nearly circular hole of radius
$r_{hole} = 1.7^{\prime\prime} = 120$ parsecs
in the H$\alpha$+[NII] image that is clearly visible in Figure 6 just
NE of the galactic center.
At the right in Figure 6 is an X-ray 
image taken from the {\it Chandra} archives.
Superimposed on the X-ray image are approximate contours showing 
the radio jets as observed 
at 1.4 GHz by Stanger \& Warwick (1986) 
and at 5 GHz by Birkinshaw \& Davies (1985). 
The (steep spectrum) double radio jets are about 1 kpc long, 
oriented at position angle $\sim45^o$. 
It is remarkable that the circular hole 
seen by the optical observers is also 
visible in the {\it Chandra} image. 
While the X-ray emitting gas (paradoxically) appears to have 
been disturbed very little by the radio lobes,  
the energy source that created the circular cavity to the 
NE has does not contain a significant 
non-thermal radio source of its own.

Assuming that the circular feature is formed by a point explosion, 
the energy required to form a hole of radius $r_{hole}$ 
in the hot gas 
is $E \approx (3P/2\rho)\cdot \rho V_{hole} = 2 \pi r_{hole}^3 P$, 
where $P = 1.9nkT$ is the local gas pressure. 
Using central values of the hot gas density $n$ 
and temperature $T$ from 
Allen et al. (2006), we estimate that the local gas pressure 
near the circular cavity 
is $P \approx 6.14 \times 10^{-10}$ dynes cm$^{-2}$. 
The energy required to form this hole is therefore 
$E \approx 200 \times 10^{51}$ ergs. 
This energy far exceeds that of a 
single supernova, so the cavity was probably formed with energy 
released near the central black hole. 
Since the chaotic warm gas velocities observed 
by Caon, Macchetto \& Pastoriza (2000) near the hole are 
$v \sim \pm 50-100$ km s$^{-1}$, the age of the hole cannot exceed 
$t_{hole} \approx 2r_{hole}/v \approx 2 \times 10^6/v_{100}$ yrs 
where $v_{100}$ is the 
local warm gas velocity in units of 100 km s$^{-1}$.
Evidently, the central AGN-black hole 
in NGC 4636 has been active during the last few million years. 

\section{Transient Activity in the Hot Gas}

Unlike most other similar galaxy/groups, 
{\it XMM} X-ray spectra of the hot gas in the NGC 5044 group 
show that the hot gas within about 30 kpc 
from the central galaxy 
does not have a uniform temperature at each radius 
(Figure 5; Buote et al. 2003).
In most galaxy groups and clusters 
the hot gas temperature rises from the center and peaks at 
about 0.1 of the virial radius, then decreases further out.
The emission-weighted (single) temperature profile 
in NGC 5044 has a similar thermal structure. 
However, {\it XMM} spectra reveal that 
NGC 5044 contains an additional component 
of hotter gas in which the temperature 
{\it decreases} outward from the 
center (Buote et al. 2003) 
as shown by the long-dashed line in the top panel of Figure 5.
Since this hotter (and therefore less dense) gas 
in NGC 5044 must be buoyant, it is natural to associate 
it with the dust-transporting regions discussed in 
the previous section.
The ``multi-phase'' character of the X-ray spectrum 
of NGC 5044 is also evident from the hot gas density 
irregularities visible in the {\it Chandra} image 
(Fig. 1 of Buote et al. 2003) 
with isophotes shown in our Figure 1.
It appears that the X-ray feature toward the 
south in NGC 5044 is roughly coincident with 
the H$\alpha$+[NII] and PAH extensions in that same region.
The enhanced X-ray emission in this region may 
result from slightly denser gas that is undergoing 
dust-assisted cooling in the buoyant plume.
Overall, observations indicate many dusty plumes 
in NGC 5044, but only the plume toward the south is currently 
producing warm gas out to $\sim 5$ kpc.

X-ray spectra and images of NGC 4636 also reveal evidence 
of recent, possibly somewhat more violent activity in 
this galaxy/group.
{\it Chandra} observations presented by Jones et al. (2002) 
show a symmetric 
X-shaped parabolic pattern of enhanced emission, 
presumably caused by shock waves, that enclose 
cavities several kpc to the east and west of the galactic center.
Jones el al. interpret this as the result of two 
symmetric and nearly simultaneous 
point explosions ($\sim 10^{57}$ ergs) 
offset from the galactic center 
that occurred about $3 \times 10^6$ yrs ago. 
No radio emission has been observed in 
these cavities which are somewhat misaligned from 
the current radio jets shown in Figure 6 
(Birkinshaw \& Davies 1985; Stanger \& Warwick 1986).

The X-ray observations of NGC 4636 
and their explanation do not necessarily require a 
(buoyant) mass 
outflow that could create the extended 
dust emission observed at 70$\mu$m.
However, strong evidence for outflows in NGC 4636 is provided 
by the {\it XMM} spectra 
and images shown by O'Sullivan, Vrtilek and Kempner (2005).
Their abundance map of NGC 4636 shows a patch 
of gas with high abundance about 25-30 kpc toward the 
southwest of the galactic core. 
O'Sullivan et al. interpret this as evidence of an older 
(collimated, plume-like) gas outflow 
that transported gas from the galactic center that was  
enriched by Type Ia supernovae.
Consequently,
there is now good X-ray evidence that the same kind of transient 
collimated outflow has occurred in both of these galaxies 
having extended far-infrared emission.

\section{Final Remarks}

We discuss two optically normal elliptical
galaxies, NGC 5044 and NGC 4636, in which we 
discovered spatially extended interstellar 
dust at 70$\mu$m.
In NGC 5044 
we observe asymmetrically extended interstellar 
8$\mu$m PAH emission associated with warm gas and 
extending about 5 kpc from the center of the galaxy. 
In view of the short sputtering lifetime for extended dust, 
$\sim 10^7$ yrs, and the commonplace optical appearance 
of both galaxies, 
we conclude that the extended dust observed  
cannot result from a recent merger with a gas-rich galaxy.
The stellar morphologies, spectra, colors and old ages 
of these two galaxies
appear to be normal with no evidence of an 
additional merger component.
Instead, we propose that the extended excess 
dust is created by stellar mass loss 
in the central $\sim1$ kpc and stored 
in small disks or clouds in the galactic cores.   
These disks are intermittently disrupted and 
heated by energy released by accretion onto the 
central black holes (AGN feedback). 
Finally, the heated dusty gas is buoyantly 
transported out to several kpc where it is observed.
The physical extent of the extended dust is governed by the 
buoyant velocity and either the grain sputtering time 
or the dust-assisted cooling time, whichever is shorter. 

X-ray observations of both NGC 5044 and 4636 show unusual, 
transient features consistent with recent 
energy releases in or near the galactic cores. 
Both galaxies show disorganized, highly fragmented 
optically visible dust clouds 
in the central 100-200 parsecs that are also highly transient. 
Finally, the timescales for all these processes are comparable 
to the sputtering lifetime of the extended dust,
about $10^7$ yrs. 

We are particularly privileged to view NGC 5044 during 
a brief interval when asymmetric, 
radially extended 8$\mu$m PAH and H$\alpha$+[NII] emission 
is present in the outflowing gas. 
We explain this remarkable observation as a result of 
dust-assisted cooling in 
a buoyant plume of hot dusty gas. 
We show with a simple calculation that dust can cool 
buoyant gas to $\sim 10^4$ K which emits the optical 
emission lines observed. 
The warm gas phase is maintained in 
thermal equilibrium near $\sim 10^4$ K  
by radiative losses and UV heating from post-AGB stars. 
Based on the collective Balmer line luminosities of 
normal elliptical galaxies, we conclude that 
this warm gas can only last for $\sim 10^5$ yrs 
before it thermally melts into the dominant hot 
interstellar phase. 
%and this accounts for the rarity of extended warm gas 
%features as in NGC 5044.
Finally, the buoyant outflow described here is 
an essential feature of flows that 
successfully resolve the cooling flow problem 
with the outward transport of both mass and energy 
from the galactic core to large radii, i.e. a 
circulation flow (Mathews, Brighenti \& Buote 2004).

\vskip.1in
\acknowledgements
This work is based on observations made with the Spitzer Space
Telescope, which is operated by the Jet Propulsion Laboratory,
California Institute of Technology, under NASA contract 1407.
Support for this work was provided by NASA through Spitzer
Guest Observer grant RSA 1276023.
Studies of the evolution of hot gas in elliptical galaxies
at UC Santa Cruz are supported by
NASA grants NAG 5-8409 \& ATP02-0122-0079 and NSF grant
AST-0098351 for which we are very grateful.

\clearpage
\begin{figure}%1
\figurenum{1}
\centering
\vskip2.in
\includegraphics[bb=50 166 422 619,scale=0.8,angle=90]{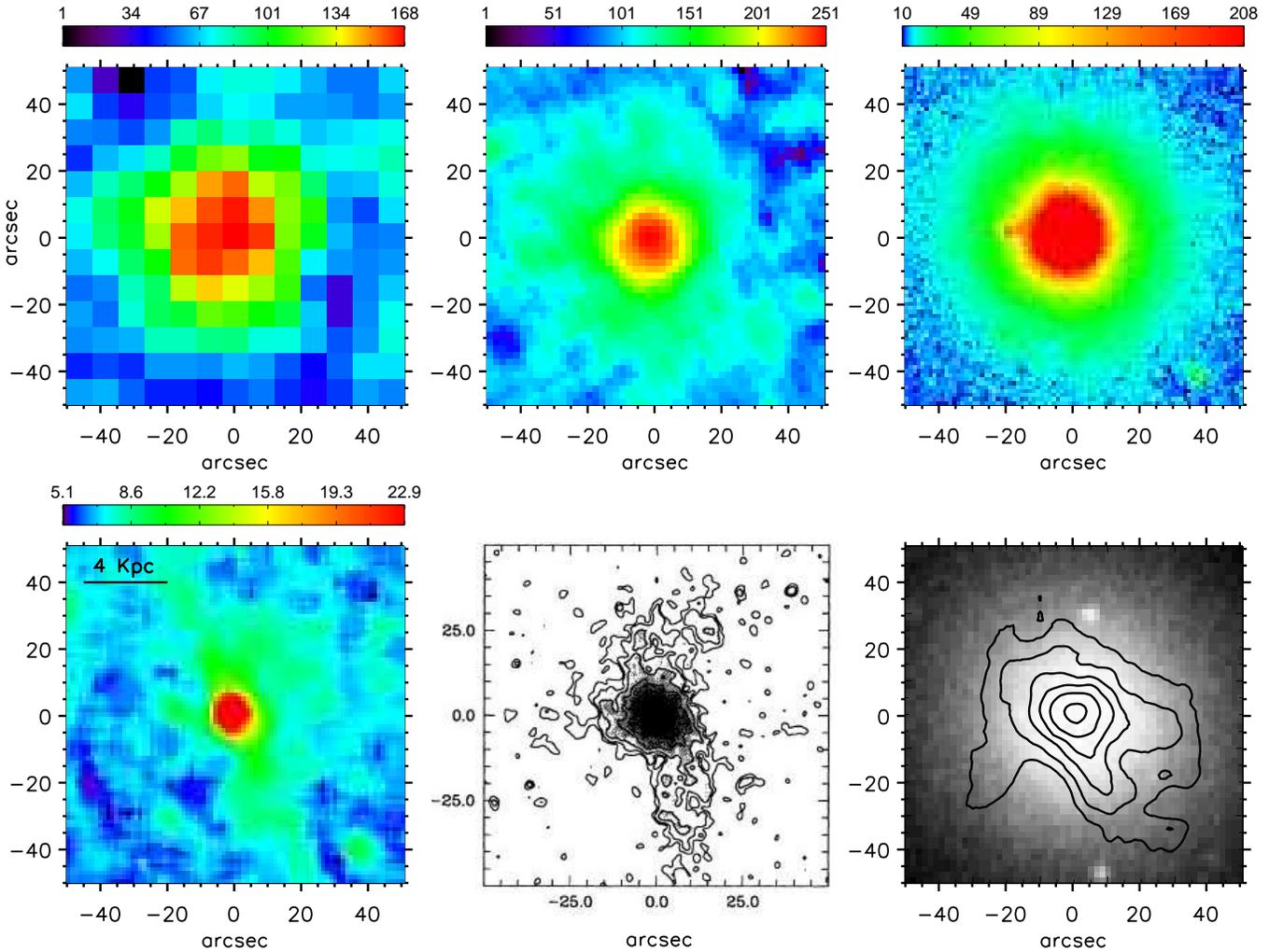}
%\vskip.7in
\caption{Comparison of optical, infrared and X-ray emission 
from NGC 5044. 
The physical scale is 160 parsecs/${\prime\prime}$.
The top three panels from left to right
show {\it Spitzer} images 
at (1) 160 $\mu$m, (2) 70 $\mu$m, 
and (3) 8 $\mu$m.
A small source about
20$^{\prime\prime}$ to the East of the center of NGC 5044,
is revealed in the 8 $\mu$m image. This source is
coincident with a point-like source in the $B-I$ image of 
Goudfrooij et al. (1994) and appears as a background galaxy 
in the HST image.
The second row of panels from left to right shows: 
(1) the difference image 8-4.5 $\mu$m, 
(2) isophotes of H$\alpha$+[NII] emission 
from warm gas in NGC 5044 taken from Goudfrooij et al. (1994) and 
(3) isophotes from the {\it Chandra} 
X-ray images superimposed on an optical image from the 
Digital Sky Survey. 
Surface brightness values in the color bars are presented in units
of $\mu Jy/arcsec^2$.
}
\label{f1}
\end{figure}

\clearpage
\begin{figure}%2
\figurenum{2}
\centering
\vskip2.in
\includegraphics[bb=50 166 422 619,scale=0.8,angle=90]{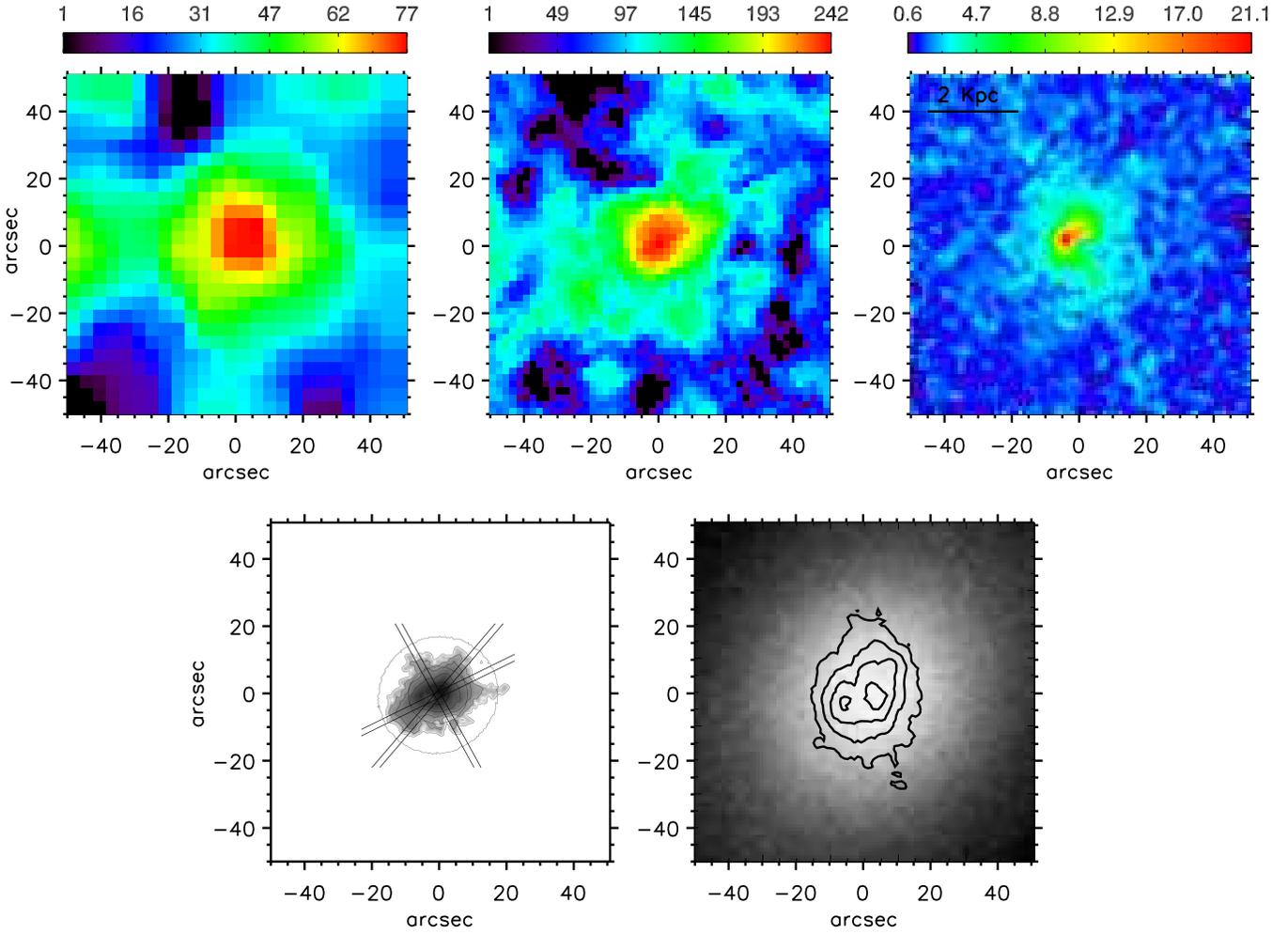}
%\vskip.7in
\caption{Comparison of optical, infrared and X-ray emission
from NGC 4636.
The physical scale is 73 parsecs/${\prime\prime}$.
The top three panels from left to right show {\it Spitzer} images
at (1) 160 $\mu$m, (2) 70 $\mu$m, 
and (3) the difference image 8-4.5 $\mu$m.
Surface brightness values in the color bars are presented in units
of $\mu Jy/arcsec^2$.
The second row of panels from left to right 
shows: (1) isophotes of H$\alpha$+[NII] emission
from warm gas in NGC 4636 taken from Caon et al. (2000) and 
(2) isophotes from the {\it Chandra}
X-ray images superimposed on an optical image from the
Digital Sky Survey.
}
\label{f2}
\end{figure}

\clearpage
\begin{figure}%3
\figurenum{3}
\centering
%\vskip2.in
%%\includegraphics[bb=90 216 522 569,scale=0.9,angle= 270]
%\includegraphics[bb=90 166 522 519,scale=1.0,angle= 0]
\includegraphics[bb=50 236 422 669]{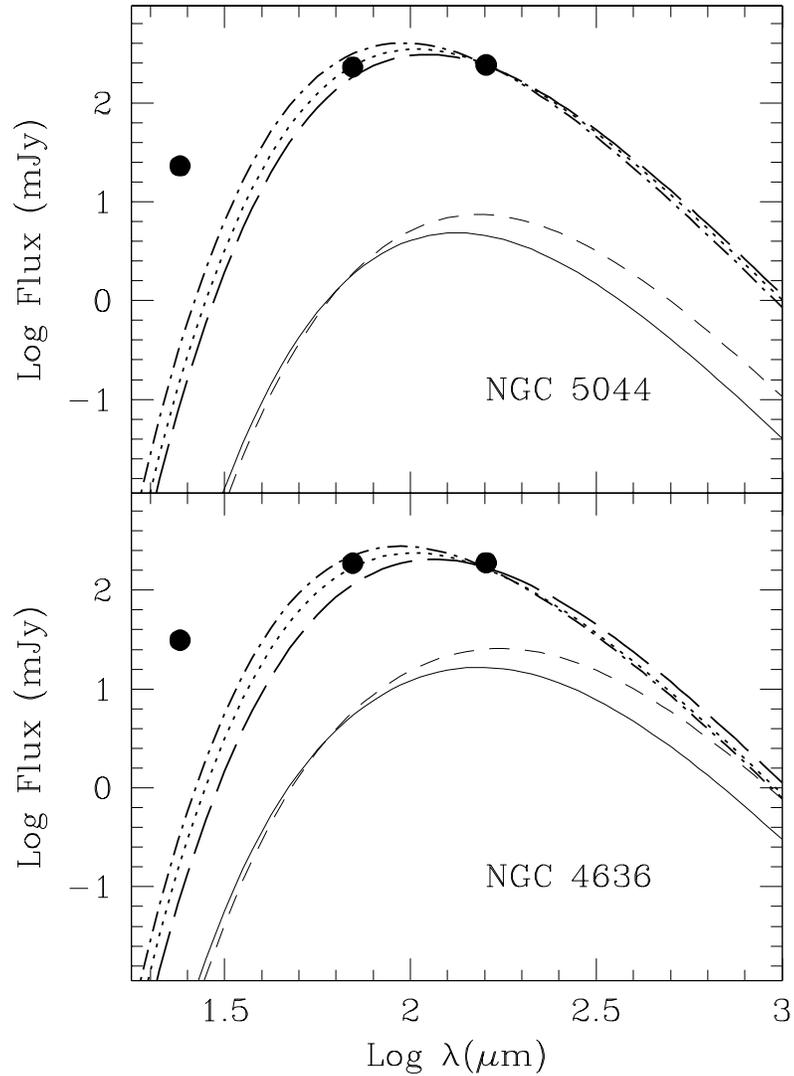}
%\vskip.7in
\caption{
Solid circles show {\it Spitzer} MIPS flux observations 
for NGC 5044 and 4638 at 24, 70 and 160 $\mu$m.
The lower two thin curves show the far-infrared 
steady state dust SED expected 
from normal stellar mass loss balanced 
by sputtering destruction in the hot gas. 
The short-dashed and solid lines show the 
SED based on initial grain size distributions with maxima 
at $a_{max} = 1$ and 0.3 $\mu$m respectively.
These models are clearly insufficient to explain the 
observations of either galaxy -- an additional source 
of extended dust is required.
Each panel shows several models of the dust SED with extra dust, 
designed to fit the interstellar 70 and 160 $\mu$m 
fluxes (but not 24 $\mu$m which is circumstellar).
As explained in the text, these models are characterized 
by two parameters $(f_{\alpha},r_{ex})$ which have the 
following values:
{\it Upper panel:} (300,5kpc) ({\it dotted line}), 
(165,10kpc) ({\it long-dashed line}), 
(800,2kpc) ({\it dash-dotted line}); 
{\it Lower panel:} (60,4kpc) ({\it dotted line}),
(26,10kpc) ({\it long-dashed line}), 
(133,2kpc) ({\it dash-dotted line}).
}
\label{f3}
\end{figure}

\clearpage
\begin{figure}%4
\figurenum{4}
\centering
\vskip2.in
\includegraphics{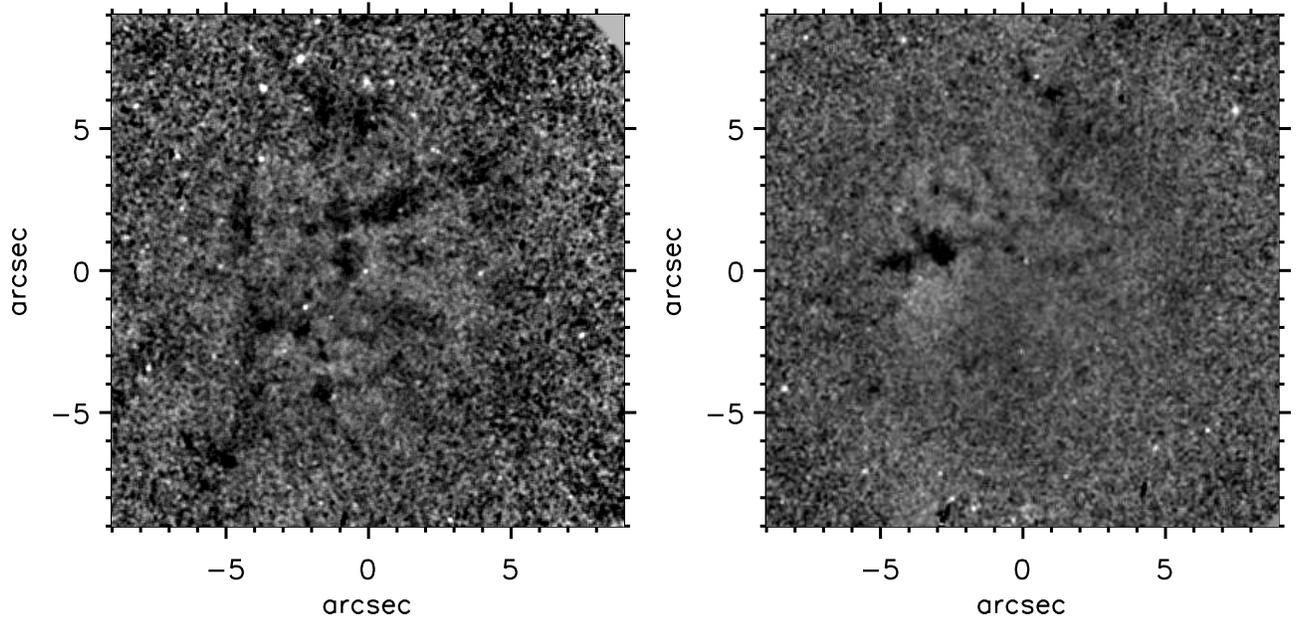}
%\vskip.7in
\caption{ {\it Hubble Space Telescope} images of the 
central $\sim 1-2$ kpc in NGC 5044 (left panel) 
and 4636 (right panel).
The optical dust maps were generated from archival HST data
recorded with the Wide Field and Planetary
Camera 2 (WFPC2) using the broad-band filters F791W and F547M.
We first modeled the stellar contribution, recorded in the
broad--band image, with elliptical isophotes using a two pass
procedure to properly mask out and remove bad pixels and
background/foreground objects. 
The dust map
was then generated from the ratio of the continuum image over
its purely stellar isophotal model. 
Dust absorption features are evident as
dark filamentary structures in the maps.
}
\label{f4}
\end{figure}

\clearpage
\begin{figure}[ht]%5
\figurenum{5}
\centering
%\vskip2.in
%%\includegraphics[bb=90 216 522 569,scale=0.9,angle= 270]
%\includegraphics[bb=90 166 522 519,scale=1.0,angle= 0]
\vskip2.in
%orig bb is BoundingBox: 22 16 596 784
%\includegraphics[bb=70 166 422 669,angle=-90]
%\includegraphics[bb=192 236 536 554,angle=-90,scale=0.8]
%\includegraphics[bb=62 256 526 554,angle=-90]
\includegraphics[bb=62 256 526 554,angle=-90]
%[bb=50 236 422 669]22 236 596 554
{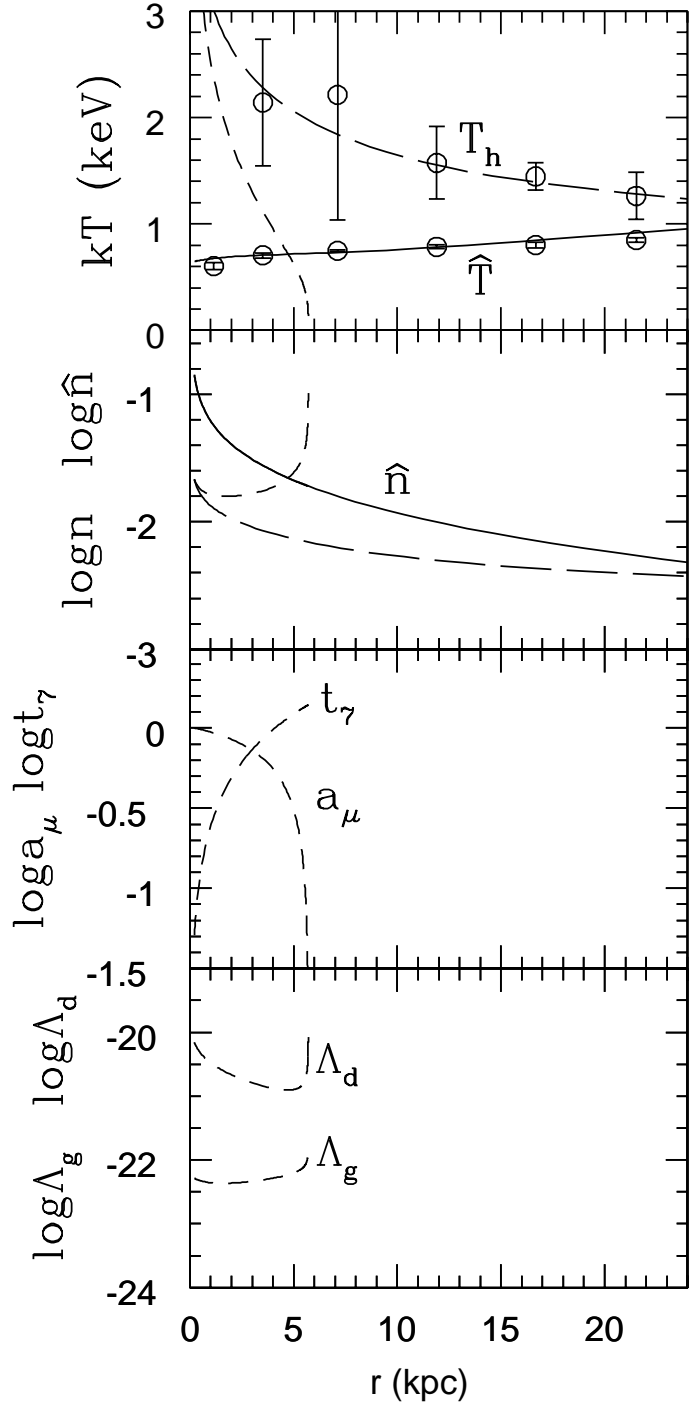}
\vskip.7in
\caption{
Approximate radial evolution of dust cooling in the 
constant-velocity buoyant outflow in NGC 5044.
{\it Top panel:} temperature (K) of a dust-free buoyant element 
(long-dashed line) and dust-filled buoyant element 
(short-dashed line) compared to that of the ambient gas (solid line),
{\it Second panel:} electron density (cm$^{-3}$) of the buoyant element
(long-dashed line) compared to that of the ambient gas (solid line) 
and that in dusty buoyant gas (short-dashed line),
{\it Third panel:} time $t_7$ in $10^7$ yrs for the buoyant element
to reach each radius (short-dashed line), 
radius $a_{\mu}$ of the grains in microns (short-dashed line),
{\it Bottom panel:} radiative cooling coefficient 
$\Lambda_{gas}$ (lower short-dashed line) 
and dust-assisted cooling coefficient
$\Lambda_{dust}$ (upper short-dashed line), both in erg cm$^{3}$ s$^{-1}$.
}
\label{f5}
\end{figure}

\clearpage
\begin{figure}%6
\figurenum{6}
\centering
\vskip2.in
\includegraphics[angle=0]
{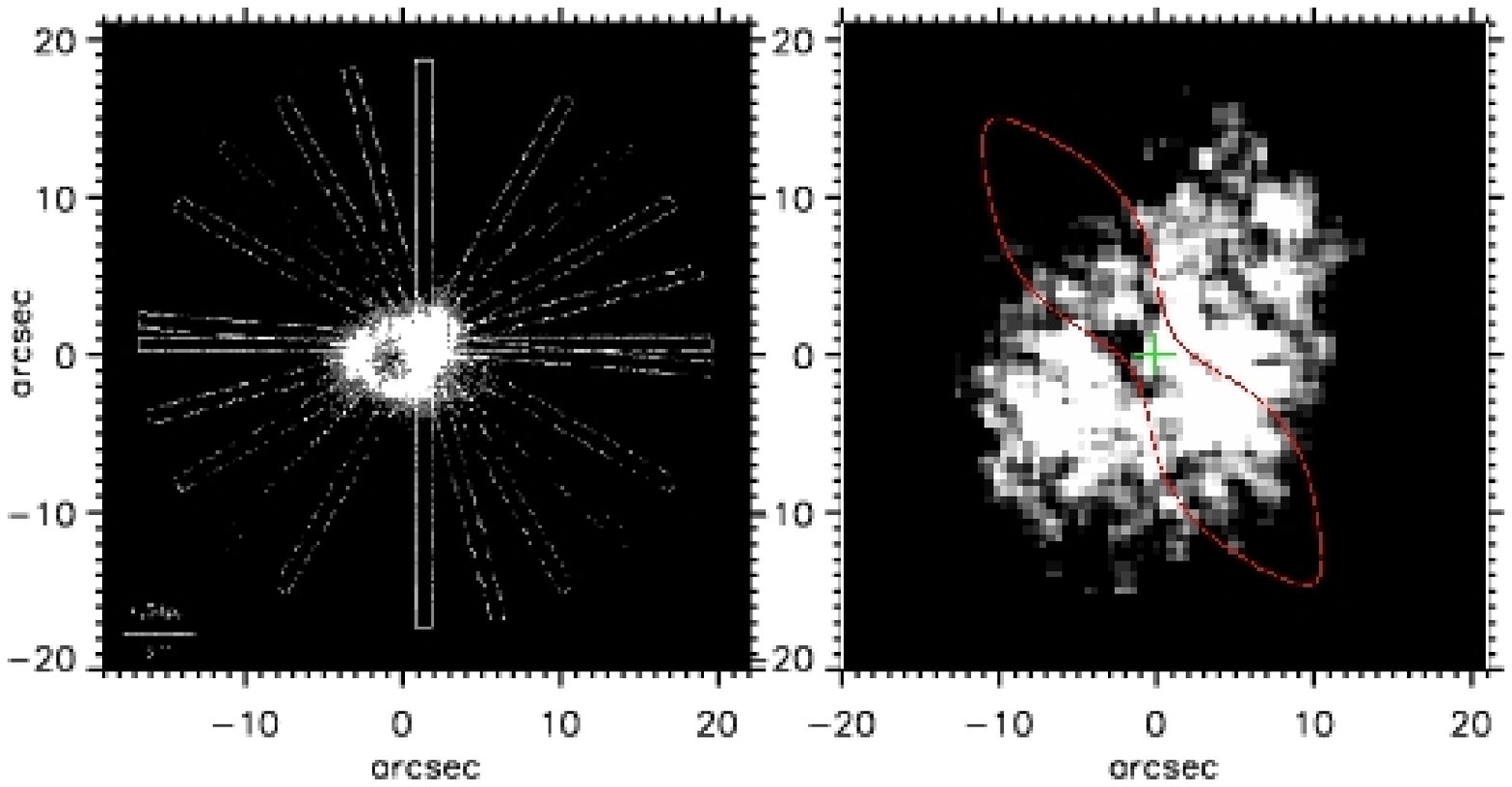}
%\vskip.7in
\caption{
{\it Left Panel:} An H$\alpha$+[NII] image of NGC 4636 
from Zeilinger et al. (1996) also showing the orientation 
of their slit spectra. These authors note that the slits 
are centered about 1.5$^{\prime\prime}$ from the galactic center.
{\it Right Panel:}
{\it Chandra} X-ray image from the {\it Chandra} 
archives with approximate radio contours of the double jet
based on Birkinshaw \& Davies (1985) and Stanger \& Warwick (1986).
The contours are symmetric about 
the green cross that shows the position of the galactic center 
based on a compromise between images from the Digital Sky Survey 
and the centroid of the HST stellar image.
}
\label{f6}
\end{figure}


\begin{references}

\reference{} Allen, S. W., Dunn, R. H., Fabian, A. C., 
Taylor, G. B. \& Reynolds, C. S. 2006, MNRAS, 272, 21

\reference{} Annibali, F., Bressan, A., Rampazzo, R., 
Zeilinger, W. W. \& Danese, L. 2006, A\&A (in press)
(astro-ph/0609175)

\reference{} Arimoto, N., Matsushita, K., Ishimaru, Y.,
Ohashi, T. \& Renzini, A. 1997, ApJ, 477, 128

\reference{} Birkinshaw, M. \& Davies, R. 1985, 
ApJ, 291, 32

%ages of E gals from IR spectra
\reference{} Bregman, J. N., Temi, P. \& 
Bregman, J. D. 2006, ApJ, 647, 265

%unusual PAH emission from NGC 4697
\reference{} Bregman, J. D., Bregman, J. N. \& Temi, P. 2006,
to appear in {\it The Spitzer Science Center 2005 
Conference: Infrared Diagnostics of Galaxy Evolution},
(astro-ph/0604369)

%\reference{} Bressan, A. et al. to appear in 
%{\it The Spitzer Science Center 2005
%Conference: Infrared Diagnostics of Galaxy Evolution},
%(astro-ph/0604068)

%Spitzer IRS spectra of Virgo early-type galaxies; silicate em.
\reference{} Bressan, A. et al. 2006, ApJ, 639, L55

\reference{} Buote, D. A., Lewis, A. D., Brighenti, F. \&
Mathews, W. G. 2003, ApJ, 594, 741

\reference{} Buson, L. M. et al. 1993, A\&A, 280, 409

\reference{} Caon, N., Macchetto, D. \& Pastoriza, M. 2000,
ApJS, 127, 39

\reference{} Demoulin-Ulrich, M. H., Butcher, H. R. \& 
Boksenberg, A. 1984, ApJ, 285, 527

\reference{} Draine, B. T. \& Anderson, N. 1985, 
ApJ, 292, 494

\reference{} Fazio, G. G. et al. 2004, ApJS, 154, 10

\reference{} Goudfrooij, P., Hansen, L, Jorgensen, H. E., 
\& Norgaard-Nielsen, H. U. 1994, A\&AS, 105, 341

\reference{} Hopkins, P. F., Narayan, R., Hernquist, L., 
2006, ApJ, 643, 641

\reference{} Jones, C., Forman, W., Vikhlinin, A., 
Markevitch, M., David, L., Warmflash, A., Murray, S., 
\& Nulsen, P. E. J. 202, ApJ, 567, L115

\reference{} Krishna Kumar, D. \& Thonnard, N. 
1983, AJ, 88, 260

\reference{} Lauer, T., R. et al. 2005, AJ, 129, 2138

\reference{} Loewenstein, M., et al. 2001, ApJ, 555, L21

\reference{} O'Sullivan, E., Vrtilk, J. M. \& Kempner, 
J. C. 2005, ApJ, 624, L77


\reference{} Mathews, W. G. 1989, AJ, 97, 42

\reference{} Mathews, W. G. \& Brighenti, F. 2003, 
ApJ, 590, L5

%formation of low mass stars in CFs
\reference{} Mathews, W. G. \& Brighenti, F. 1999, 
ApJ, 526, 114

%time-dept circulation flows
\reference{} Mathews, W. G., Brighenti, F. \& 
Buote, D. A. 2004, ApJ, 615, 662

%\reference{} Nagar, N. M., Falcke, H., Wilson, A. S. \&
%Ho, L. C. 2000, ApJ, 542, 186

\reference{} Phillips, J. P. \& Ramos-Larios, G. 2006, 
MNRAS, 368, 1773

\reference{} Piovan, L, Tantalo, R.,\& Chiosi, C. 2003, 
A\&A, 408, 559

\reference{} Poulain, P. 1988, A\&AS, 72, 215
\reference{} Poulain, P. \& Nieto, J.-L. 1994, A\&AS, 103, 573

\reference{} Ravindranath, S. et al. 2001, AJ, 122, 653

\reference{} Stanger, V. \& Warwick, R. 1986, MNRAS, 220, 363

\reference{} Sutherland, R. S. \& Dopita, M. A. 1993, 
ApJS, 88, 253

\reference{} Temi, P., Brighenti, F., \& Mathews, W. G. 2007,
ApJ (in press) (astro-ph/0701431)

\reference{} Temi, P., Mathews, W. G., Brighenti, F., \& 
Bregman, J. D. 2003, ApJ, 585, L121

\reference{} Tonry, J. et al. 2001, ApJ, 546, 681

%\reference{} Tsai, J. C. \& Mathews, W. G. 1996, ApJ, 
%468, 571

\reference{} Tsai, J. C. \& Mathews, W. G. 1995, ApJ,
448, 84

\reference{} Whittle, M., Rosario, D. J., Silverman, J. D., 
Nelson, C. H. \& Wilson, A. S. 2005, ApJ, 129, 104

\reference{} Zeilinger, W. W. et al. 1996, A\&AS, 120, 257

\end{references}
\end{document}